
\magnification=1200
\baselineskip=18truept

\line{\hfill RU--95--49}
\vskip 2truecm
\centerline{\bf The overlap is not a waveguide.}

\vskip 1truecm
\centerline{Rajamani Narayanan$^{a}$ and Herbert Neuberger$^{b}$}
\vskip 0.5truecm
\centerline{${}^a$\it School of Natural Sciences,
Institute for Advanced Study,}
\centerline{\it Olden Lane, Princeton, NJ 08540}
\vskip .5truecm
\centerline{${}^b$\it Department of Physics and Astronomy,
Rutgers University }
\centerline{\it Piscataway, NJ 08855-0849}
\vfill
\centerline{\bf Abstract}
\vskip 0.75truecm
Golterman and Shamir falsely claim that a waveguide
model modified by adding many charged bosonic spinors, in
the limit of an infinite number of matter fields, becomes
identical to the overlap
if in the target theory every fermion appears in four
copies. Their modified model would give wrong results
even in the vectorial four flavor massless Schwinger model,
while a dynamical simulation of this model with the overlap
works correctly. In this note we pinpoint the error in the
derivation of Golterman and Shamir.
\vfill
\eject
In a recent letter [1] the overlap method of defining chiral gauge theories
was criticized on the grounds that it is similar to a Yukawa model known
as the ``waveguide''. In the seventh paragraph of [1] the main result is
identified as an exact equivalence between the overlap of ref. [2] and
a ``modified waveguide model'' with Higgs hopping
parameter $\kappa$ set to 0 and Yukawa coupling $y$ set to 1. The equivalence
is allegedly shown for a situation where there are 4$n$ identical
chiral families. The proof of the equivalence concludes with the
last equation in [1] (eq. (42)) which is claimed to have been rigorously
established.

The modified waveguide differs in a major way from the original waveguide.
The discussion of the properties of the modified waveguide is highly
speculative. However, we don't need to engage in imprecise arguments here
since the main result of ref. [1] is false. The error made by Golterman and
Shamir was specified as a possible pitfall in our paper [3] (last
paragraph in section 6 there), also listed in reference [3] of [1].

The modified waveguide, when analyzed using the second quantized
transfer matrices introduced in [2],
$T_{\pm} (U)$, depending parametrically on the background gauge field
$U$, leads after the integration of all matter fields to:
$$
\left [ {{Tr \left [ T_-^L (1) T_-^L (U) T_+^L (U) T_+^L (1)\right ]}\over
{{\sqrt { Tr \left [ T_-^{2L} (1) T_-^{2L} (U) \right ]}{\sqrt
{ Tr \left [ T_+^{2L} (1) T_+^{2L} (U)\right ] }}}}} \right ]^{4n}.\eqno (*)$$
$L$ is a large integer. The original waveguide model gives an effective
action equal to the logarithm of the numerator of the above expression.
It is claimed in ref. [1] that as $L\to\infty$  $(*)$ converges to the overlap.
This is incorrect in general because the ground states of $T_- (U)$ and
$T_+ (U)$ will be orthogonal to each other for large chunks of gauge
field space. Such gauge field backgrounds will be generated with
finite probability in a Monte Carlo simulation using $(*)$.
In particular this happens when the gauge field background
is close to a continuum connection on a principal bundle carrying nonzero
instanton number. The overlap gives a vanishing result reflecting the
known zero modes while $(*)$ would typically approach some finite limit, the
traces being dominated by a combination of second
quantized fermionic ground and excited states.

Let us explain this in some more detail: The transfer matrices are
exponents of bilinears in fermion creation/annihilation operators
that conserve a total fermion number. When $L\to\infty$ the three
traces in $(*)$ will be dominated by specific states of definite
fermion number. (Typically, states with the same fermion number will
have nonzero inner products and exact accidental degeneracies will not
happen.) Suppose the background has lattice topological charge 1.
The single way $(*)$ could vanish in the limit $L\to\infty$
is when the fermion numbers dominating the two traces in the
denominator of $(*)$ are different. However, deforming the
gauge background towards a configuration of zero topological
charge and where we have near
degeneracy in $T_+ (U)$ (the single matrix that is substantially
sensitive to the topology of the background [3]) we see that
there will be situations where the fermion numbers of
the dominating states in the denominator will be the same and
$(*)$ will have a finite limit. The regions in gauge field space
where the overlap is strictly zero and $(*)$ is non--zero are
of codimension zero and there is no obvious mechanism to
suppress them with probability one.

The claimed equivalence could not have been correct {\sl a priori}
since for any finite $L$ the modified waveguide has exact
global symmetries that are known to be violated in the continuum.
This observation holds also in the context of a vector-like theory, showing
that the failure of the modified waveguide is independent of the chiral
nature of the fermionic content of the target theory.
More precisely, had we replaced
the fermion subroutines implementing the overlap in our Monte Carlo
simulation of the four flavor massless Schwinger model by
subroutines implementing the
modified waveguide, we would have obtained on a torus
$<\prod_{f=1}^4 \bar\psi_f \psi_f>=0$.
Numerical results obtained with the overlap yield a
nonvanishing $<\prod_{f=1}^4 \bar\psi_f \psi_f>$ in quantitative agreement
with continuum [4].

The unwanted preservation of global symmetries is a common feature of the
majority of the attempts to regulate nonperturbatively gauge theories
with massless fermions
using Yukawa models. For this reason, an understanding of the detailed
dynamics of these models (in particular for $F_{\mu\nu}\equiv 0$, the focus
of most recent research) is likely to be irrelevant to the problem of
putting {\sl chiral} gauge theories on the lattice. Of course, this opinion
is open to debate.

\bigskip\bigskip
\centerline{\bf Acknowledgments}
\bigskip
This research was supported in part
by the DOE under grants DE-FG02-90ER40542 (RN) and DE-FG05-90ER40559 (HN).
\bigskip\bigskip
\centerline{\bf References}
\bigskip
\item{[1]} M. F. L. Golterman, Y. Shamir, Phys. Lett. 353B (1995) 84.
\item{[2]} R. Narayanan, H. Neuberger, Nucl. Phys. B 412 (1994) 574.
\item{[3]} R. Narayanan, H. Neuberger, Nucl. Phys. B 443 (1995) 305,
(hep-th/9411108).
\item{[4]} R. Narayanan, H. Neuberger, P. Vranas, Phys. Lett. 353B (1995) 507.
\item{} I. Sachs, A. Wipf, Helv. Phys. Acta 65 (1992) 652.
\item{} R. Narayanan, H. Neuberger, P. Vranas, in preparation.

\vfill
\eject

\end